\documentstyle[12pt,epsf]{article}
\newcommand\pubdate{August, 1998}

\def\Title#1{\begin{center} {\Large #1 } \end{center}}
\def\Author#1{\begin{center}{ \sc #1} \end{center}}
\def\Address#1{\begin{center}{ \it #1} \end{center}}

\newcommand\pubblock{\rightline{\begin{tabular}{l}
         \pubdate  \end{tabular}}}
\newenvironment{Abstract}{\begin{quotation} \begin{center}
                       ABSTRACT
     \end{center}\bigskip  }{\end{quotation}}
\def\beq{\begin{equation}}
\def\eeq#1{\label{#1}\end{equation}}
\def\eeqn{\end{equation}}
\def\beqa{\begin{eqnarray}}
\def\eeqa#1{\label{#1}\end{eqnarray}}
\def\eeqan{\end{eqnarray}}
\def\CR{\nonumber \\ }
\def\leqn#1{(\ref{#1})}
\def\Acknowledgements{\bigskip  \bigskip \begin{center} 
\begin{large}
             \bf ACKNOWLEDGEMENTS \end{large}\end{center}}

\begin{document}
\begin{titlepage}
\pubblock

\vfill
\Title{Scalar Black Holes}
\vfill
\Author{Erast B. Gliner\footnote{Visiting physicist at 
Stanford Linear Acceleration Center, 
Stanford University, California 94309, USA}}
\Address{661 Faxon Ave., San Francisco, CA 94112\,{}\footnote{E-mail:
Erast@Netscape.net}}
\vfill
\begin{Abstract}

A class of exact regular spherically symmetric solutions to the 
Einstein equation obeying Dymnikova's definition 
$T_1^1=T_0^0$ of 
vacuumlike state is considered. These solutions, that may be 
interpreted as {\it black holes}, are not only 
singularity free, but also 
 do not set us thinking about the loss of information under 
gravitational collapse. However, according to the singularity 
theorems, the geometries introduced by
 these solutions inevitably 
have some causal pathology.

If vacuumlike state is reinterpreted to be a sort of 
{\it confinement}
representing a particular phase of matter, this pathology under
certain conditions does not, however, involve actual causality
violation. By means of that, the mentioned 
above class of solutions,
named in what follows {\it scalar black holes}, 
and probably much
broader variety of solutions including dynamic ones, may be
incorporated in General Relativity. There
are listed other phenomena, such as `hidden mass', 
that could help to identify the presence of vacuumlike phase.
\end {Abstract}
\vfill
\centerline{PACS numbers: 04.20.Jb; 97.60.Lf; 98.80.Bp}
 \vfill
\end{titlepage}
\def\thefootnote{\fnsymbol{footnote}}
\setcounter{footnote}{0}

\section{Introduction}

During the last decade, the internal structure of black holes has
attracted considerable attention, with issues of mass-energy
singularity and nonunitary black hole dynamics 
being in focus (e.g.,
\cite{one,two}). The latest development has produced the
perturbative quantum models of {\it string} black holes with
quantum microstates being countable, and 
closed-string Hawking
emission going on in a unitary way. The emission 
rate agrees with
the semiclassical Hawking prediction. 
quantum models still
contain the `central' singularity but in a `civilized' 
form: though the
sources of fields propagating in the black hole interior are viewed as
located at singularity, the modes of the fields 
are well defined up to
 event horizon.

Despite the fact that quantum models 
still have some shortcomings
(e.g., only integer moded states are 
accountable with the current
perturbative technique), their very existence 
shows that the loss of
information is not a must for black holes.

The confirmation of this remarkable 
discovery demands, however,
the construction of a macroscopic black 
hole model with similar
properties because a fundamental criterion of applicability of a
theory still is the preservation of macroscopic causality. The
violation of just this `principle of 
macroscopic censorship' prompted
J. Wheeler to name gravitational collapse `the greatest crises in
physics of all time' \cite{three}.

Put very simply, by their structure the 
standard black hole models
(e.g. the Schwarzschild one) provide no capacity for storage of
propagation modes and quantum numbers, which therefore are
condemned to die in gravitational trap. The 
most efforts to eliminate
this shortcoming (e.g.,\cite{four,five}) introduce the storage
capacity by proposing some quantum phenomena that collect
information in the {\it vacuum} dynamic 
region of a black hole.
Instead, as its starting point, the present 
work takes the Dymnikova
model \cite{six,seven,eight} based on the exact 
regular solution to
the Einstein equation \cite{eight,nine} obeying the condition
\beq
T_1^1=T_0^0\,,\quad \,T_2^2=T_3^3\,,
\quad \,T_k^i=0\quad if\quad i\ne k .
\eeq{oneone}

We shall consider a class of solutions 
obeying \leqn{oneone}. Each
of these solutions with final, but large 
enough, Schwarzschild mass
depicts asymptotically flat spacetime with 
an event horizon. The
presence of the horizon identifies 
them as black holes. Inside the
event horizon, there exists a {\it nonsingular mass-energy
distribution} that seems to be a natural carrier of the information,
which supposedly disappears in Schwarzschild black hole. Thus, the
imperative to admit the nonunitary development of gravitational
collapse is eliminated.

The singularity theorems \cite{ten,eleven,twelve}, which are an
integral part of GR, impose, however, severe restrictions on the use
of the regular black hole solutions: particles, following paths
allowed by such solutions, unavoidably run into some causality
violation.

The essence of difficulties is as follows. Let a nonsingular spherical
{\it core} replace the Schwarzschild singularity. By symmetry, the
metric at the core center is Minkowskian. Thus, in some vicinity of
the center, a particle can move both toward the center and outward.
As it turns out, a particle, freely moving outward, leaves the core,
enters the dynamic region, and then traverses the outer space by
crossing the event horizon. In the dynamic region, the movement of
this particle proceeds, however, against the time arrow established
by the inward falling particles, which cross the horizon from the
outer space. Thus, the regular black hole has an `oncoming time
topology' preventing consistent definition of the arrow of time in the
dynamic region. According to the third Tipler theorem \cite{twelve}
this discordance is generic rather than conditioned by a high level of
symmetry.

The widely accepted paradigm, considering geometry to be the
warrant of causal evolution of universe, implies that in these
circumstances one must reject either regular black hole solutions or
causality (if not General Relativity itself). This paradigm, however,
is not a part of GR, and looks suspicious. Indeed, the Einstein
equation
 \beq
G_{ik}=-\,\kappa \,T_{ik}
\eeq{onetwo}
{\it algebraically} links two quantities, which observationally are
mutually independent. Thus, the Einstein equation is an {\it
equivalence relation} (invariant, but otherwise just like $E=mc^2$).
Therefore properties attributed to matter should be taken on equal
terms as those related to geometry. Thus, in some circumstances just
the properties of matter can protect causality whereas the geometry
by itself {\it allows} causality violation (cf. \cite{thirteen}).

This modifies the view of gravitational collapse. A causal geometric
deficiency may be admitted in conjunction with the states of matter,
which properties prevent the actual violation of causality on their
own.

 Following this program, we examine, which properties of the matter
involved let us incorporate the regular black hole solutions in GR.
The answer amounts to three main conditions: (1) the matter is in
{\it vacuumlike state}; (2) the vacuumlike state is a {\it particle
confinement} of a kind; and (3) the high-density vacuumlike
continuum exhibits {\it combined stress pattern}.

\section{Vacuumlike continuum}

Incorporating the cosmological term, $\lambda \,g_{ik}$, in his
equation, Einstein \cite{fourteen} was the first who introduced a
Lorentz-invariant, and in this sense vacuumlike, mass-energy
distribution. He, however, assumed the term to be related to the
background of spacetime geometry. With coming the dynamic
models pioneered by Friedmann \cite{fifteen}, 
Einstein threw this term
out, but since then the idea of modified or modifiable vacuum
surfaced repeatedly.

Along with metric tensor $g_{ik}$ , the scalar-tensor theory of
gravitation \cite{sixteen,seventeen} introduces also scalar field
$\varphi $. Without the gravity-connected terms, the Lagrangian of
scalar-tensor theory reduces to the Lagrangian for massless scalar
field:
\beq
L={1 \over 2}\,g^{ab}\varphi _{,a}\varphi _{,b}-\varepsilon \left(
\varphi \right)\,,\quad \varepsilon \left( \varphi \right)\ge 0,
\eeq{twoone}
The stress-energy tensor of its homogeneous state $\left( {\varphi
_{,\,\,i}=0,\;\;i=0,\,1,\,2,\,3} \right)$ is
\beq
T_k^i=\delta _k^i\,\varepsilon \,,\quad \varepsilon =const\ge 0 .
\eeq{twotwo} .
Possessing the structure of cosmological term, this tensor is Lorentz
invariant. In this sense, the scalar field $\varphi $ is
vacuumlike. Shortly after the scalar-tensor theory 
appeared, it has been understood 
\cite{eighteen} that GR allows the existence of {\it
vacuumlike continuum} that, in distinction to the cosmological term,
represents a {\it phase of matter}. i.e., can participate in mutual
phase transitions with phases of matter having stress-energy tensors
of general algebraic type.

According to \leqn{twotwo}, for vacuumlike phase
\beq
T_0^0=T_1^1=T_2^2=T_3^3 ,
\eeq{twothree}
i.e., its energy density $\varepsilon \equiv T_0^0$ and the main
pressures $p_i\equiv -\,T_i^i$, $ i=1,2,3$ ({\it no summation}), are
related by
\beq
p_i=-\varepsilon \,,\;\quad i=1,\,2,\,3\; .
\eeq{twofour}
Thus, the energy being positive, the pressures are negative. This
seems to involve spontaneous contraction -- mechanical collapse of
continuum. In general relativity, however, this is not the case, and
\leqn{twotwo} corresponds to the equilibrium de Sitter solution with
{\it negative pressures counterbalanced by the divergence of
geodesics}, i.e., by the {\it repulsive} gravitational action. Hence,
the vacuumlike phase has no classical counterpart, and is {\it
essentially relativistic} substance comprehensible only in GR frame.

As was immediately realized, the vacuumlike phase is a candidate
for the upper state of matter attainable at growing density. Sakharov
\cite{nineteen} proposed that it is the initial cosmological state
providing nonsingular onset of expanding universe. At the same
time, there was offered nonsingular Friedmann cosmology 
\cite{twenty}. Much later development brought the vacuumlike
phase down to GUT and inflationary cosmologies
\cite{twentyone,twentytwo}.

As applied to the black hole physics, a series of works was initiated
by Markov \cite{twentythree} who proposed that quantum
corrections bound the spacetime curvature by Planckian value; this
value being reached, the effective stress-energy tensor takes the form
\leqn{twotwo}. This hypothesis, elaborated by Frolov, Markov, and
Muchanov \cite{twentyfour}, has been further studied by Poisson
and Israel \cite{twentyfive}, Balbinot and Poisson \cite{twentysix},
and Balbinot \cite{twentyseven}. In distinction from cosmology,
where \leqn{twothree} can be easily incorporated into existing
formalisms, in case of black hole the Israel junction conditions
\cite{twentyeight,twentynine} require the existence of a transition
layer between the core and vacuum. The layer cannot be vacuumlike
in the sense of (4). Also, conceived as a thin shelf, it represents a {\it
spacelike} surface \cite{twentyseven}. In other words, unlike the
timelike surface of ordinary body, it exists at a single instant of time
\cite{twentyseven}.

This brief survey outlines the sphere, which the present work
belongs to. It has been partly initiated by the essential Dymnikova's
idea \cite{six} to reject equation \leqn{twothree} as the exhaustive
macroscopic definition of vacuumlike phase. She offers equations
\leqn{oneone}, which introduce the {\it minimal spherical symmetry}
instead of homogeneity. She reasonably maintains that the medium
obeying \leqn{oneone} {\it has no unique comoving reference
frame}, so that it is a vacuumlike one. Her black hole model is based
on the exact {\it regular} solution to the Einstein equation:
\beqa
&\quad& \quad ds^2=\left( {1-{{R_g\left( r \right)} \over r}}
 \right)\,dt^2-\,\,\left( {1-{{R_g\left( r \right)} \over r}} \right)^{-1}
dr^2-r^2d\Omega ^2,\CR &\quad&d\Omega ^2\equiv d\theta ^2+
\sin ^2\theta \,d\varphi ^2\;,\quad R_g\left( r \right)\equiv r_g
\left( {1-e^{-{{r^3} \over {r_0^2r_g}}}} \right)\,,\;
\quad \,r_0\equiv \sqrt {{3 \over {8\pi \varepsilon _0}}}
\eeqa{twofive}
where $r_g$ is gravitational radius and $\varepsilon _0$ the energy
density at the center of the black hole configuration. The entire
family of regular solutions defined by the Dymnikova condition
\leqn{oneone} is considered in the next Section.

Let us consider the properties of continuum obeying homogeneous
conditions \leqn{twothree}.

Generally, the stress-energy tensor, $T_k^i$ , defines the proper
mass-energy density, $\varepsilon$, as well as the local velocity,
$u^i$, of a physical continuum, by means of the eigen-value
equation
\beq
T_a^i\,u^a=\varepsilon \,u^i .
\eeq{twosix}
If the algebraic structure of $T_k^i$ provides the vector field $u^i$
to exist, and is timelike and unique, we will name the continuum {\it
particlelike}. In this case the field $u^i$ generates a unique
congruence of paths, to which it is tangent. The {\it test particles}
(or {\it observers}, or {\it fluid elements}) propagating along
these paths form {\it comoving reference frame}. By definition,
continuum is {\it at rest} relative to its comoving frame.

Eq. \leqn{twosix} may have no solution, as, e.g., in case of free
electromagnetic field. If continuum obeys \leqn{twothree}, Eq.
\leqn{twosix} degenerates into a trivial identity 0 = 0, and the notion
of local velocity ceases to be applicable at all, just as it is for
ordinary vacuum. This is the case of homogeneous vacuumlike
continuum. Its properties, immediately evident from Eq.
\leqn{twothree}, are as follows:

1. If \leqn{twothree} holds, any vector $u^i$ satisfies \leqn{twosix},
so that the notion of the velocity of a particle relative to the
continuum loses any definite meaning.

2. In analogy with ordinary vacuum, in each reference frame free
falling in the continuum, the local physics, including the results of
the measurements over the continuum, is the same.

3. In default of the unique comoving reference frames, the concept
of Eilerian (moving with the medium) element of vacuumlike
continuum cannot be established and, by means of that, also {\it the
concept of particles that constitute} this continuum.

4. In the continuum obeying \leqn{twothree} everywhere, the 
particle creation is forbidden because of the uncertainty in the
impulse of the particles that would be created. This means that the phase transition of vacuumlike continuum to a phase with particles is
to be initiated by particlelike seeds.

5. Since the notion of a particle velocity relative to vacuumlike
continuum cannot be established, any interactions between a particle
and this continuum are independent of the particle velocity
(principle of relativity).

6. This means that there is no basis for the kinetic thermal exchange
between vacuumlike and particlelike continua. Thus, the concept of
the temperature {\it in Boltzmann's sense} cannot be introduced for
vacuumlike continuum, and by means of that, the concept of
Boltzmann's entropy. Thus, the macroscopic internal state of
vacuumlike continuum is defined by the 
{\it only parameter} -- its mass-energy density.

7. Since for particlelike phase Boltzmann's 
temperature is well defined, the 
possibility of phase transition between particlelike and vacuumlike 
phases makes us to attribute an effective temperature to 
vacuumlike phase, e.g., the temperature of the particlelike phase 
newly created in a reversible process. In distinction from the 
classical temperature, this effective temperature is not {\it an 
independent variable}, but a function of vacuumlike mass-energy 
density (cf. Eq. \leqn{twoeight} below and \cite{thirty}).

8. Because of the divergence of geodesics inside vacuumlike 
 continuum, two test particles, free falling on neighboring 
geodesics, are accelerated one relative to another. 
This {\it gravitational repulsion} is balanced by the 
compressing negative pressure \leqn{twofour}, 
so that the continuum is in equilibrium.

A simple model of the phase transition of a (not necessary
homogeneous initially) particlelike medium to a homogeneous
vacuumlike state is delivered by the relativistic 
equation of perfect fluid motion:
\beq
\left( {\varepsilon +p} \right)\,{{\delta \,u^i} \over {\delta \,s}}=
p_{,\,a}\,\left( {g^{ia}-u^iu^a}
\right)\;,\quad \;\varepsilon \ge 0 .
\eeq{twoseven}
Let us assume that, with growing density $\varepsilon $,
 an internal attraction (negative pressure) appears in the fluid. 
Then the fluid will undergo self-compression.  But when the pressure 
$p\to -\varepsilon $, its gradient, $p_{,\,a}$,
vanishes. Such a self-aligning process could 
produce a homogeneous state with pressure 
satisfying \leqn{twofour}. The final state is
Lorentz invariant and therefore vacuumlike; actual particles
disappear as {\it observable} entities, and the particlelike structure
is virtual.

A simple example may be useful to overview 
the distinction between actual and virtual behaviors. 
According to Gibbons and Hawking
\cite{thirtytwo}, an observer, falling freely in 
de Sitter spacetime, reveals the presence
 of thermal radiation of temperature
\beq
T =\sqrt {{2 \over {3\pi }}}\,{{\hbar c} \over {k_B}}\,
\sqrt {G\varepsilon }\quad \left( K \right) .
\eeq{twoeight}
When actual radiation is present, any two observers in relative
motion feel it differently, whereas the value \leqn{twoeight} is
observer-independent. This means that the Gibbons-Hawking
radiation is virtual, rather than actual. From this point of view,
\leqn{twoeight} indicates not the real presence of radiation but 
only determines the equilibrium conditions between continuum 
and a measuring device: in equilibrium, the device has 
Gibbons-Hawking temperature. As a part of the device state, 
the radiation captured by the device from the vacuumlike 
thermal bath is, of course, real. Just as we may expect,
the equilibrium conditions do not depend on the velocity of the
device and defined by the only parameter -- the mass-energy 
density of continuum 

Generalizing this example, we learn first that the
 existence of phase transitions between vacuumlike 
and particlelike phases is the {\it
requirement of thermodynamics}; and, second, that this 
transition is conditioned by the presence of 
{\it inhomogeneity} breaking off the
ban on the particle creation in homogeneous vacuumlike phase.

Thus, we come to the view of the phase transition to
vacuumlike state as brought on by the rise in the particlelike phase
of internal attraction becoming the dominant internal force. 
Particles, confined inside the final bound state, can display its
presence only within the limits of the uncertainty principle, as 
the {\it virtual} particle structure. 

Along with vacuum, the resulting phase resembles the 
phenomena of particle physics in two ways. First, it is a kind of 
confinement \cite{thirtyone} similar to the quark confinement where an
 attraction between quarks conceals their properties as free 
particles. Second, the overwhelming attractive forces appear at high 
concentration of mass-energy. This 
resemblance lets us anticipate that, having much
in common with the known in particle physics, the vacuumlike
confinement does not violate quantum conservation laws.

A profound approach that covers the aspects mentioned above 
follows from the fact that the characteristic features of vacuumlike 
continuum are closely similar to those of ordinary vacuum except 
two distinctions: the non-vanishing Lorentz-invariant stress-energy 
tensor and the proposed ability to retain quantum charges. These 
distinctions are just the ones required by classical and quantum 
conservation laws for considering vacuum a phase of matter. If one 
is governed by the idea of similarity of vacuumlike phase to the 
ordinary vacuum, the values of quantities describing a vacuumlike 
phase should be considered {\it zero-point values} from which the 
corresponding values for particlelike phase should be reckoned. 

This suggests that the phase transition to vacuumlike state may be 
considered a {\it shift} of the {\it zero-point} densities of  mass-
energy  and quantum charges from their zero values (which are 
characteristic of ordinary vacuum) to non-zero levels, what provides 
the accommodation of the energy and quantum charges of the 
particles eliminated as actual physical entities. From this point of 
view, vacuumlike continuum is vacuum with {\it shifted} zero-point 
levels. This approach seems to be the most consistent though it is, 
probably, not absolutely necessary for understanding the black hole 
models considered.

Considering vacuumlike continuum in this way, we, supposedly, 
should admit that inside continuum the spectrum of particle masses 
can change. For our goals, it is enough to accept the intuitive 
expectation that the mass-energy density, naturally connected with 
an actual particle inside the continuum, should exceed the 
mass-energy density of vacuumlike phase (which we considered above to 
be zero-point level). Then, some light-weight particles may be 
nonexistent in vacuumlike continuum. We will return to this 
suggestion in Section 4. 

Now let us turn our attention to a localized and therefore
inhomogeneous distribution of vacuumlike phase. 
According to the
Einstein equation, the externally observed mass of a localized object 
depends only on the internal mass-energy 
distribution. Therefore, its vacuumlike properties have not 
come to light in its external behavior
as a mechanical and gravitating body. The internal equilibrium,
however, essentially depends on the algebraic structure of its 
stress-energy tensor. One faces two main possibilities: either 
vacuumlike continuum exhibits only purely 
normal stress (as perfect fluid) or
combined stress (as, e.g., elastic body). In the former case no
regular steady-state inhomogeneous solutions exist. Thus, it is our
guess that, at least at collapse-induced-densities, the vacuumlike
phase endures {\it combined stress}.

Under combined stress, the stress-energy tensor ceases to be Lorentz 
invariant exactly. This, however, 
does not ruin the vacuumlike
behavior rooted in the lacking of actual particles or , from a more 
general point of view, in the shift of the zero-point values. All the 
properties
listed above, therefore, remain valid, 
at least at the scale small in
comparison with the characteristic length 
of inhomogeneity when
the mass-energy distribution must be considered 
as the united whole.

\section{Scalar black holes}

We shall restrict our consideration by the Dymnikova family of
solutions to the Einstein equation with diagonal stress-energy tensor
obeying the Dymnikova condition \leqn{oneone}. In this case 
metric can be written in the form
\beq
ds^2=A(r)\,dt^2-{{dr^2} \over {A(r)\,}
}-r^2\,d\Omega ^2 ,
\eeq{threeone}
and vice versa, from \leqn{threeone} follows \leqn{oneone}.

 If, for a given mass-energy distribution $\varepsilon =\varepsilon
\,\left( r \right)$, the Schwarzschild mass of the whole configuration
\beq
M={{4\pi } \over {c^2}}\,\int\limits_0^\infty {\varepsilon \,\zeta
^2d\varsigma <\infty } ,
\eeq{threetwo}
the solution to the Einstein equation is given by
\beq
A\left( r \right)=1-{{2G} \over {c^4}}\,
{{4\pi } \over r}\,\int\limits_0^r {\varepsilon \,
\varsigma ^2d\varsigma } .
\eeq{threethree}
Thus, within the restriction \leqn{threetwo}, the solution is defined
by the mass-energy distribution $\varepsilon =\varepsilon \,\left( r
\right)$ , which can be chosen arbitrary. The stress-energy tensor is
then defined as follows:
\beq
T_0^0=T_1^1=\varepsilon \,,\;\quad T_2^2=
T_3^3=\varepsilon +{r \over 2}\varepsilon _{,\,r} .
\eeq{threefour}
Note, $T_2^2=T_3^3<0$ corresponds to positive pressure. The 
latter may be assumed unacceptable for
 vacuumlike phase . According to \leqn{threefour}, to eliminate 
positive pressure, the mass-energy density must
drop not faster than $r^{-2}$ everywhere, except some vicinity of
$r=0$.

Let us introduce dimensionless quantities:
\beq
\chi \equiv {r \over {r_g}}\,\;\quad and\;\quad \mu
\equiv {\varepsilon \over {\varepsilon _m}}\,\,\,;\quad
\quad r_g\equiv {{2GM} \over {c^2}}\,,\quad \;\
{\varepsilon _m}\equiv {1 \over {4\pi }}\,{{Mc^2} \over {r_g^3}} .
\eeq{threefive}
For each {\it arbitrary chosen} smooth $\mu =\mu \left( \chi \right)$,
the function\footnote{The use of notation `$A\left( \chi \right)$'
that is not correct in strong mathematical sense should not lead to
misunderstanding.}
\beq
A\left( \chi \right)=1-{1 \over \chi }\,\int\limits_0^\chi
\mu \varsigma ^2d\varsigma
\eeq{threesix}
determines a subclass of similar spacetimes distinguished by the
value of the mass {\it M}. Substituting  $k\ mu \left( \chi \right)$
for $\mu \left( \chi \right)$ , we, for each given $\mu \left( \chi
\right)$, can find such critical value of $k=k_{cr}$ that $A\left( \chi
\right)$ is non-negative for $k\le k_{cr}$, but for $k>k_{cr}$
changes its sign at some $\chi =\chi _h$. Since $A\left( 0
\right)=A\left( \infty \right)=1$, the zeros $\chi _h$ are emerging in
pairs. Thus, there are two essentially different classes of solutions.
One, with $A\left( \chi \right)$ being everywhere non-negative,
contains {\it starlike} solutions. Another one, distinguishing by the
presence of zeros of $A\left( \chi \right)$, represents {\it black hole} solutions. As shown in the next Section, these solutions are
physically acceptable only if in some vicinity of $\chi =0$ the matter
is in vacuumlike phase; otherwise causality cannot be preserved. 
We will name these solutions {\it scalar black holes}.

The Dymnikova Eq. \leqn{twofive} gives an example of solutions of 
both classes: a starlike solution for mass-energy density 
$\varepsilon_0$ being small enough, and 
a scalar black hole solution for larger
$\varepsilon _0$ values.

The surfaces where $A\left( \chi \right)$ changes sign represent
horizons of a kind, similar, but not identical to the Schwarzschild
horizon. A scalar black hole has at least a {\it pair of horizons}. We
will only consider black holes with one pair of horizons and distinct
the {\it inner horizon}, $\chi =\chi _i$, and the {\it outer} one, $\chi
=\chi _e$. On the horizons, the $\chi ,\,t$ -coordinates are mutually
converted from spacelike to timelike or vice versa.\footnote{This 
fact, of cause, does not interfere with using
these coordinates for analytic description.} In the {\it dynamic
region} between horizons, no particle can be at rest relative to $\chi
$ which is timelike there, whereas {\it t} is spacelike. In the region $\chi_i>\chi >0$, in the {\it core}, the orientation of $\chi ,\,t$ -
coordinates is the same as in 
the {\it outer space} $\chi \,\,>\chi _e$ :
$\chi $ is spacelike and {\it t} timelike. By symmetry, the metric at
the center $\chi =0$ of the core is Minkowskian; it is closed to the
de Sitter one in its vicinity. Since $\mu \left( \chi \right)$ is an
arbitrary function, the configuration of scalar black holes can have a 
variety of patterns with horizons that can be located inside or outside 
the sphere $r=r_g$.

\section{Challenge to causality}

The structure of a scalar black hole, dominated by the presence of
 pairs of horizons, is radically distinct from the Schwarzschild black
hole. Two phenomena, the {\it oncoming time topology} and {\it
quasistatic mass-energy distribution} in the dynamic region,
challenge the very understanding of causality in GR.

To a certain extent, scalar black hole recalls the Schwarzschild one.
A {\it `testing observer'}, radially falling onto this black hole from
the outer space, reaches the outer horizon in a finite proper time and
enters black hole interior. From this moment, in analogy with the
Schwarzschild case, he perceives his journey into diminishing {\it r} 
as changes in the environment in 
the course of time - until, again in a final proper time, he has 
found himself on the inner horizon (instead 
of singularity). Recalling the argumentation addressed to the
singularity in Schwarzschild black hole, one might erroneously
conclude that, like the Schwarzschild singularity, the core lies in the 
absolute future of the outer horizon. 
With observer's deepening into
the core, the Schwarzschild-like scenario is, however, over.

Free falling into or out of a scalar black hole \leqn{threeone} is
determined by equations:
\beq
r_{,\,t}^2=c^2A^2\,\left( {1-{A \over b}} \right)\,;\quad
r_{,\,t\,t}^{}=c^2A\,A_{,\,r}\,\left( {1-{3 \over 2}
\,{A \over b}} \right) ,\eeq{fourone}
\beq
r_{,s}^2=b-A\,,\;\;b=const\;;\quad r_{,s\,s}=
-{1 \over 2}\,A_{,r} .CR
\eeq{fourtwo}

Eqs. \leqn{fourtwo} define the velocity, $r_{,\,t}$, and the
acceleration, $r_{,\,t\,t}$, of an observer. Since $A=0$ on the
horizons, a test body approaching the inner horizon from the core
becomes unvisible from the inside. A similar effect 
is detected by an outer observer keeping his eye on 
a body falling on the outer horizon 
(well known black hole feature). As shown below, these horizons
cannot be, nevertheless considered as one-way membranes.

Velocity, $r_{,\,s}$, and acceleration, $r_{,\,ss}$, in units of the
proper time of a free falling observer are given by (18). The general
pattern is as follows. If the integral of motion $b<1$, the equation
$A\left( r \right)- b=0$ has two zeros
{\it r}=$\mathord{\buildrel{\lower3pt\hbox
{$\scriptscriptstyle\rightharpoonup$}}\over r} $ and {\it r}=$\mathord{\buildrel{\lower3pt\hbox
{$\scriptscriptstyle\leftharpoonup$}}\over r} $. One, say 
$\mathord{\buildrel{\lower3pt\hbox{$\scriptscriptstyle
\rightharpoonup$}}\over r} $, lies in the outer space, another,
$\mathord{\buildrel{\lower3pt\hbox{$\scriptscriptstyle
$}}\over r} $, {\it inside the core.} At
$\mathord{\buildrel{\lower3pt\hbox{$\scriptscriptstyle
\rightharpoonup$}}\over r} $ and
$\mathord{\buildrel{\lower3pt\hbox{$\scriptscriptstyle
\leftharpoonup$}}\over r} $ observer's velocity is vanishing. At
$\mathord{\buildrel{\lower3pt\hbox{$\scriptscriptstyle
\rightharpoonup$}}\over r} $ he is accelerated toward the black hole 
(quasi-Newtonian attraction), but inside the core, where 
$A_{,\,r}<0$,
outward. Thus, from 
{\it r}=$\mathord{\buildrel{\lower3pt\hbox{$\scriptscriptstyle
\rightharpoonup$}}\over r} $ he begins his fall 
into black hole, slows down inside the core, stops at {\it r}
=$\mathord{\buildrel{\lower3pt\hbox{$\scriptscriptstyle
\leftharpoonup$}}\over r} $, accelerates outward, and returns to {\it
r}=$\mathord{\buildrel{\lower3pt
\hbox{$\scriptscriptstyle\rightharpoonup$}}\over r} $ for 
a final proper time. Then, he repeats this
cycle. If $b>1$, the observer flies through the black hole and
escapes to infinity. If $b=1$, he gets an unstable rest at the core
center; any disturbance pushes him back toward outer space. The
problem is: crossing the dynamic region on his way to the core, the
observer travels in the direction of time, which is opposite 
to the one on his way back.

Thus, on geometric basis, the arrow of time in the dynamic region
between horizons cannot be chosen uniquely, this is a {\it NOT-
region} (not orientable in time). Using the term coined apparently 
by Wald \cite{eleven}, scalar black hole has {\it oncoming time
topology}. By means of that its geometry cannot secure the 
course of physical events from causality violation.

Such geometries customary are considered unacceptable. However,
Riemannian geometry is alien to the notion of direction to be a
distinctive geometric feature. The geometry as a bearer or a guard of
the arrow of time cannot therefore be a primary 
notion and is only justified as a synonymic to some 
non-geometric circumstances (cf. Sec. 1).

The arrow of time is usually viewed to be connected with
irreversible cause-and-effect sequence of events in particlelike
phase. This view cannot be translated into vacuumlike phase
 because a sequence of causally connected 
events represents a unique
reference frame that cannot come to existence inside vacuumlike 
phase. Deeper insight on the reason for this provides the uncertainty 
principle, which restricts the life-span of virtual particles just 
enough to make it insufficient to pass or acquire definite
information and by means of that let virtual particles be converted to 
actual ones.

The inability of vacuumlike phase to distribute information clearly
manifests itself in the absence of sound. According to
\leqn{twofour}, the velocity of sound in vacuumlike phase
$c_s=c\,\left( d\varepsilon /dp \right)^{{1/2}}$
has an imaginary value.

Thus, for the core to influence the arrow of time in the dynamic
region, it must contain an internal carrier of the time arrow, i.e.,
particlelike phase with developed cause-and-
effect relations between
particles. If no particles exist inside the core, would they
be particles created there or penetrated from the outside, the source
of oncoming time topology is lacking.

In the absence of particles serving as a reference system, the
spontaneous particle creation in vacuumlike phase is forbidden. 
Another way for a particle to appear in the core is its penetration 
from the dynamic region. Since the interaction between particle and
vacuumlike phase does not depend on particle's velocity, a particle 
inside vacuumlike phase is in the state of free falling and can leave 
the core. But as we discussed it 
in Sec. 2, a changeover from ordinary
vacuum to vacuumlike phase can change the particle identity,
primarily its mass. Say, only the excess of effective density of
particle over the vacuumlike phase density contributes to 
observed particle's mass, so that in the limit of equal 
densities a particle is captured by the phase.

Our guess is: for a particle to exist inside a vacuumlike phase, 
the mass density associated with particle 
(say $\rho _{pl}\sim
c^3m_{pl}^4/\hbar ^3$ with $m_{pl}$ being the particle mass) 
must exceed that of vacuumlike phase. 
Then, vacuumlike phase is free of
particles if its mass-energy density is large enough and the spectrum 
of particles is bounded from above (what is now commonly 
accepted). If the top quark with mass 
$\sim 150\,m_p$ is considered
to be the heaviest particle, a vacuumlike continuum is free of
particlelike phase if its mass density is larger than
\beq
c^3\,\hbar ^{-3}\,\left( {150m_p} \right)^4\
\sim  10^{26}\,g\,cm^{-3} .
\eeq{fourthree}
This value may be overestimated because quark does 
not exist as a free particle.

Interestingly enough, nothing makes us bring Planck density 
in. This implies that the scalar black hole 
theory is in no need for quantum
gravity; GR and the tools of the standard particle physics 
seem to be sufficient for the further development.

Thus, if the density inside the core is not less than some critical
density, e.g. \leqn{fourthree}, there is no emission from the core, 
 and the arrow 
of time in the dynamic region is established by matter falling from
the outside. Actual bodies {\it fall down} into such black 
hole, rather than {\it `fall up'} (in contrast to the 
fictitious `testing observer').

This way of establishing the arrow of time is seemingly indeed
realized in the scalar black holes originating in gravitational
collapse if the bulk transition to vacuumlike phase takes place at the
critical or a larger density. Then the formation of vacuumlike
core begins not earlier than the critical density is attained, and,
proceeding layer by layer, keeps the density of each layer being not
less than the critical one. Having been formed, the core is
completely filled with vacuumlike phase, because its boundary -- the 
inner horizon -- is immersed in vacuumlike phase. Indeed, according
to \leqn{threethree}, the distribution $\varepsilon =\varepsilon
\,\left( r \right)$ must continue beyond the inner horizon into the
dynamic region, otherwise the horizons, and the black hole 
itself, do not appear.
The core, therefore, does not contain cavities 
where particles falling
from the outside could be turned 
back to the dynamic region.

Thus, we expect that breaching causality by particles moving
outward are {\it geometrically admissible}, 
but not allowed {\it by
the conditions on matter involved}, so that 
the effect of oncoming
topology is nullified \cite{thirtythree}.

Another challenging feature of scalar black holes appears just
because the mass-energy distribution $\varepsilon =\varepsilon
\,\left( r \right)$ continues beyond the inner horizon into
the dynamic region.

Reasoning naively, one may speak about an element of the 
mass-energy distributed in the dynamic region, and, 
applying \leqn{fourtwo}, find
that each such element falls on the horizon 
in a final proper time.
Should this be the case, the dynamic region would be free of 
mass-energy in a final time of any local observer.

In application to the vacuumlike phase, this, however, is
wrong. The easiest way to show that is to recall that the
notion of the local velocity is inapplicable to this phase, so
that the movement originating in any given initial velocity
distribution is unrelated to this phase. 
This implies that, the phase
cannot be dragged in the motion along paths, which the
particlelike phase follows. Thus, as soon as a vacuumlike 
mass-energy distribution obeys the Einstein equation, we should
acknowledge the possibility of its presence in the dynamic region.

We may take a look at all that from the different 
point of view. Since
the concept of vacuumlike phase 
generalizes the notion of ordinary
vacuum, we should regard the mass-energy density of vacuumlike
phase as the local {\it zero-point energy density} (Sec. 2). This
returns us to the
idea of a particle as primarily an {\it excess} in
 density above this zero-point level, with 
this excess to be the only carrier of properties of
actual particles. The excess, i.e. particlelike 
phase, is washed out
from the dynamic region, whereas the zero-point energy 
density distribution is, strictly speaking,  just the object, 
which we have named `scalar black hole'. 

Thus, our consideration of the macroscopic
aspects of the stationary scalar black hole models suggests 
that a black hole, which is free of singularity, causality
violation, and the loss of quantum information, is 
theoretically possible.

An essential shortcoming of our approach 
is the restriction by the
stationary case. This is justified by mathematical
simplicity rather than by physical reasoning. 
A realistic model of 
spherically symmetric scalar black hole may happen to incorporate
 the interaction of black hole with its vicinity not only by means of 
gravity
and Hawking emission, but also due to 
the vacuumlike phase propagation.
The features of the Dymnikova family of solutions seemingly 
supports the last possibility because vacuumlike phase in these 
solutions is present 
far from the spot where the phase transition under
collapse is expected to take place.

\section{In search of vacuumlike phase}

Since the black hole interior is unavailable for observation, let us
 make some suggestions on searching for vacuumlike phase in other 
spots. 

The phase transition to vacuumlike phase is the result a catastrophic 
self-compression, which enormously changes the magnitude of 
pressures. 
This leads us to believe that in
terms of the field theory vacuumlike phase 
corresponds to a deep
local minimum of effective potential. Then the
barrier penetration to particlelike phase may be slow enough
for the existence of long-living metastable vacuumlike formations
that may produce observable phenomena. Even the state of
the ordinary vacuum can be unattainable (just as the absolute
zero of temperature). At the `bottom of the universe' we will then find the low density vacuumlike 
phase. This phase is not required to
be homogeneous as ordinary vacuum, homogeneous by definition,
so that it is reasonable to speak about vacuumlike {\it clouds}.
They may influence the observation in several ways.

First, the clouds act gravitationally. For an {\it outside} observer, a 
cloud represents hidden mass. At the same time, the 
cosmological objects
{\it inside} a cloud undergo outward accelerations. 
These effects can misshape the pattern of homogeneous 
cosmological accelerations.

Second, as a whole, vacuumlike clouds behave as ordinary bodies, 
so that they can be in orbital movements round gravitating 
bodies or fall on them forming a halo. The latter must disturb 
the orbital movement of bodies revolving round the central body. 

Third, a vacuumlike cloud may serve as `anti-gravity' (diverging) 
lens. Distorting the background of faint galaxies, this can create the 
illusion of void.

Fourth, a vacuumlike cloud can be dense enough for 
the phase transition 
to particlelike state to create particle-antiparticle pairs. The 
subsequent particle-antiparticle annihilation will then produce
characteristic annihilation electromagnetic emission from this cloud.

Fifth, the presence of vacuumlike phase can 
influence the output of nuclear reactions by means of 
capturing light particles, e.g. neutrino. 

 Sixth, the assumed ability of vacuumlike phase to hold 
quantum charges can induce some subtle differences 
between particles and  antiparticles,
e.g., such as between $K^\circ $ and $\bar K^\circ $.

\Acknowledgements

I would like to express my gratitude to I. Dymnikova, 
W. Israel, A. Linde, M. Peskin, A. Silbergleit, 
and R. Wagoner for fruitful
suggestions and discussions. The hospitality shown to me at the
Stanford Linear Acceleration Center gave me the opportunity to
accomplish this work.

\end{document}